\documentstyle[12pt]{article}
\begin{document}
\begin{titlepage}
\bigskip
\begin{center}
Tel Aviv University Preprint TAUP 2238-95, 10 May 1995\\
Bulletin Board hep-ph@xxx.lanl.gov  - 9503289
\end{center}
\begin{center}
{\Large\bf   How to Search for\\ Doubly Charmed Baryons and Tetraquarks\\}
\end{center}
\begin{center}
{\bf      Murray A. Moinester  \\
R. \& B. Sackler Faculty of Exact Sciences,\\
School of Physics and Astronomy, \\
Tel Aviv University, 69978 Ramat Aviv, Israel,\\
E-mail: murray@tauphy.tau.ac.il\\}
\end{center}
\vspace{0.3in}
\begin{center}
Expanded Version of Workshop Contribution:\\
"Physics with Hadron Beams with a High Intensity Spectrometer",\\
Workshop Information: S. Paul, CERN PPE,\\ CH1211 Geneva 23,
Switzerland, E-mail: snp@vsnhd8.cern.ch \\
Workshop Dates: Nov. 24-25, 1994, Feb. 2-3, 1995, March 20, 1995 \\
\end{center}
\bigskip
\begin{center}{\Large\bf  Abstract}\end{center}
\normalsize
Possible experimental searches of doubly charmed baryons and tetraquarks at
fixed target experiments with high energy hadron beams and a high intensity
spectrometer are considered here. The baryons considered are:
$\Xi_{cc}^{+}$ (ccd), $\Xi_{cc}^{++}$ (ccu), and $\Omega_{cc}^{+}$ (ccs);
and the tetraquark is T ($cc\bar{u}\bar{d}$). Estimates are given of
masses, lifetimes, internal structure, production cross sections, decay
modes, branching ratios, and yields. Experimental requirements are given
for optimizing the signal and minimizing the backgrounds. The discussion is
in the spirit of an experimental and theoretical review, as part of the
planning for a CHarm Experiment with Omni-Purpose Setup (CHEOPS) at CERN.
The CHEOPS objective is to achieve a state-of-the-art very charming
experiment, in the spirit of the aims of the recent CHARM2000 workshop.
\end{titlepage}
\section*{Introduction}
The Quantum Chromodynamics hadron spectrum includes doubly charmed baryons:
$\Xi_{cc}^{+}$ (ccd), $\Xi_{cc}^{++} (ccu), $ and $\Omega_{cc}^{+}$ (ccs),
as well as ccc and ccb. Properties of ccq baryons were discussed by Bjorken
\cite {bj}, Richard \cite {ric}, Fleck and Richard \cite {fr}, Savage and
Wise \cite {sw}, Savage and Springer \cite {ss}, Kiselev et al. \cite
{kis,kis2}, Falk et al. \cite {fal}, and by Bander and Subbaraman \cite
{ban}. Singly charmed baryons are an active area of current research
\cite{app,tav,kle,pdg,sie,kor}, but there are no experimental data on the
doubly charmed variety. A dedicated double charm state of the art
experiment is feasible and required to observe and to investigate such
baryons. The required detectors and data acquisition system would need very
high rate capabilities, and therefore would also serve as a testing ground
for LHC detectors. Double charm physics is in the mainstream and part of
the natural development of QCD research. The discussion here is in the
spirit of an experimental and theoretical review, as part of the planning
for a CHarm Experiment with Omni-Purpose Setup (CHEOPS) at CERN \cite
{paul}. The CHEOPS objective is to achieve a state-of-the-art very charming
experiment, in the spirit of the aims of the recent CHARM2000 workshop
\cite {kk}.

The ccq baryons should be described in terms of a combination of
perturbative and non-perturbative QCD. For these baryons, the light q
orbits a tightly bound cc pair. The study of such configurations and their
weak decays can help to set constraints on phenomenological models of
quark-quark forces \cite{fr,ros}. Hadron structures with size scales much
less than 1/$\Lambda_{qcd}$ should be well described by perturbative QCD.
This is so, since the small size assures that $\alpha_s$ is small, and
therefore the leading term in the perturbative expansion is adequate. The
tightly bound (cc)$_{\bar{3}}$ diquark in ccq  may satisfy this condition.
For ccq, on the other hand, the radius is dominated by the low mass q, and
is therefore large. The relative (cc)-(q) structure may be described
similar to mesons $\bar{Q}q$, where the (cc) pair plays the role of the
heavy antiquark. Savage and Wise \cite {sw} discussed the ccq excitation
spectrum for the q degree of freedom (with the cc in its ground state) via
the analogy to the spectrum of $\bar{Q}q$ mesons. Fleck and Richard \cite
{fr} calculated excitation spectra and other properties of ccq baryons for
a variety of potential and bag models, which describe successfully known
hadrons. The ccq calculations contrast with ccc or ccb or b-quark physics,
which are closer to the perturbative regime. As pointed out by Bjorken
\cite{bj}, one should strive to study the ccc baryon. Its excitation
spectrum, including several narrow levels above the ground state, should be
closer to the perturbative regime. The ccq studies are a valuable prelude
to such ccc efforts.

A tetraquark ($cc\bar{u}\bar{d}$) structure (designated here by T) was
described by Richard, Tornqvist, Bander and  Subbaraman, Lipkin, Nussinov,
and Chow \cite {ric,ban,lip,tor,nus,cho}. Tetraquarks with only u,d,s
quarks have also been extensively studied \cite {bein}. The doubly charmed
tetraquark is of particular interest, as the calculations of these authors
indicate that it may be bound. Some authors \cite{ric,ban,nus,cho} compare
the tetraquark structure to that of the antibaryon $\bar{Q}\bar{u}\bar{d}$,
 which has the coupling $\bar{Q}_{\bar{3}}(\bar{u}\bar{d})_3$. In the T,
the tightly bound (cc)$_{\bar{3}}$ then plays the role of the antiquark
$\bar{Q}$. The tetraquark may also have a deuteron-like meson-meson weakly
bound $D^{*+}D^0$ component, coupled to 1$^+$, and bound by a long range
one-pion exchange potential \cite {tor,nus}. Such a structure has been
referred to as a deuson by Tornqvist \cite {tor}. Recalling that pion
exchange, in the quark picture,  corresponds to light quark exchanges, the
deuson is analogous with the H$_2$ molecule; where the heavy and light
quarks play the roles of protons and electrons, respectively. The discovery
of such an exotic hadron would have far reaching consequences for QCD, for
the concept of confinement, and for specific models of hadron structure
(lattice, string, and bag models). Detailed discussions of exotic hadron
physics can be found in recent reviews \cite {lan1}. Some other exotics
that can be investigated in CHEOPS are: Pentaquarks $uud\bar{c}s,
udd\bar{c}s, uds\bar{c}s, uud\bar{c}c, udd\bar{c}c, uds\bar{c}c$
\cite{moi}, Hybrid $q\bar{q}g$ \cite {zie}, $us\bar{d}\bar{d}$ U$^+$(3100)
\cite {u3100}, uuddss H hexaquark \cite {H}, uuddcc H$_{cc}$ hexaquark
\cite {cho}, $q\bar{q}s\bar{s}$ or $q\bar{q}g$ C(1480) \cite {lan1}, and
$\bar{c}\bar{c}qqqqq$ heptaquark \cite {ban}.  But we do not discuss these
various exotic hadrons in detail in this report.

 Should only the $cc\bar{u}\bar{d}$ ($D^{*+}D^0$) be bound; or should the
$c\bar{c}d\bar{u}$ ($D^{*-}D^0$) also be bound? The $D^{*+}D^0$ state, if
above the DD$\pi$ threshold, can only decay strongly to doubly charmed
systems. but it is easier to produce only one c$\bar{c}$ pair, as in
$D^{*-}D^0$. However, this  state has numerous open strong decay channels.
These include charmonium plus one or two pions and all the multipion states
and resonances below 3.6 GeV, and it is therefore not strong interaction
stable. A $D^{*-}D^0$ state is also unlikely to be bound. In a deuson,
bound by pion-exchange, the sign of the potential  which binds the two D
mesons depends on the product of the sign of the two vertices associated
with the pion exchange. The sign of the D$^*$ vertex depends on T$_z$, the
z-component of isospin, which changes from +1 to -1 in changing from
positive to negative D$^*$. Therefore, if the potential is attractive in
the case of $D^{*+}D^0$, it will be repulsive in the case of $D^{*-}D^0$.
Consequently, if one accepts the calculations \cite {tor,nus} for a bound
$D^{*+}D^0$, we can conclude that the $D^{*-}D^0$ is unbound. Still, in the
$D^{*+}D^0$ search, it may be of value to look at $D^{*-}D^0$ data.
Although no peak is expected, the combinatoric backgrounds may help
understand those for $D^{*+}D^0$.

\section*{Mass of ccq Baryons and T}

Bjorken \cite {bj} suggests mass ratios M($\Omega^{++}_{ccc}/\Psi$) = 1.60
and M($\Omega^{-}_{bbb}/\Upsilon$=1.57), which follows from the
extrapolation of M($\Delta^{++}/\rho,\omega$) and M($\Omega^-/\phi$). He
assumes the validity of the "equal-spacing" rule for the masses of all the
J=3/2 baryons, which gives the possibility to interpolate between ccc, bbb,
and ordinary baryons. The masses of ccq baryons with J=1/2 were estimated
relative to the central J=3/2 value. The cc diquark is a color antitriplet
with spin S=1. The spin of the third quark is either parallel (J=3/2) or
anti-parallel (J=1/2) to the diquark. The magnitude of the splitting  is in
inverse proportion to the product of the masses of the light and heavy
quarks. These are taken as 300 MeV for u and d, 450 MeV for s, 1550 MeV for
c, and 4850 MeV for b. The equal spacing rule for J=3/2, with n$_i$ the
number of quarks of a given flavor, is then \cite{bj}:\\
\begin{equation}
M=1/3[1232(n_u+n_d) + 1672 n_s + 4955 n_c +14852n_b]. \\
\end{equation}
This equation for ccq gives  results close to those of Fleck and Richard
\cite {ric,fr}, who also estimate the tetraquark mass. Fleck and Richard,
and Nussinov \cite {nus} have  shown that ccq and $cc\bar{u}\bar{d}$ masses
near 3.7 GeV are consistent with expectations from QCD mass inequalities.

The estimates lead to  masses \cite{bj,ric}: \\
(ccs), 1/2+, 3.8 GeV;       \\
(ccu), 1/2+, 3.7 GeV;          \\
(ccd), 1/2+, 3.7 GeV;          \\
$(cc\bar{u}\bar{d})$, 1/2+, 3.6 GeV;          \\

\section*{Lifetime of ccq Baryons and T.}

The $\Xi_{cc}^{++}$ and $\Omega_{cc}^{+}$ decays should probobly be
dominated by spectator diagrams \cite {bj,fr,app,lif1,lif2,lif3} with
liftimes about $~200fs$, roughly half of the $D^{0}$ or $\Xi_{c}^{+}$.
Fleck and Richard \cite {fr} suggest that positive interference will occur
between the s-quark resulting from c-decay, and the pre-existing s-quark in
$\Omega_{cc}^{+}$. Its lifetime would then be less than that of
$\Xi_{c}^{++}$. Bjorken \cite {bj} and also Fleck and Richard \cite {fr}
suggest that internal W exchange diagrams in the $\Xi_{cc}^{+}$ decay could
reduce its lifetime to around  $~100fs$, roughly half the lifetime of the
$\Lambda_{c}^{+}$. The lifetime of the T should be much shorter, according
to the pattern set by the D$^{*+}$ lifetime. These estimates are consistent
with the present understanding of charmed hadron lifetimes \cite
{app,lif1,lif2,lif3}. One expects that predominantly doubly charmed hadrons
are produced with small momentum in the center of mass of the colliding
hadrons. They are therefore sufficiently fast in the laboratory frame. The
lifetime boost in the laboratory frame for a ccq baryon is roughly \cite
{fra} $\gamma \approx \sqrt{p_{in}/2M_N}$, if it is produced at the center
of rapidity with a high energy hadron beam of momentum p$_{in}$. For a CERN
experiment with p$_{in}$ $\approx$ 400 GeV/c, this corresponds to $\gamma
\approx 15. $, with ccq energies near  55 GeV.

\section*{Production Cross Section of ccq Baryons}

One can  consider production of doubly charmed hadrons by proton and Sigma
and pion beams. Pion beams are more effective in producing high-X$_F$ D$^-$
mesons, as compared to $\Sigma^-$ beams. Here, X$_F$ designates the Feynman
X$_F$-value,  X$_F = p_D/P_{beam}$, evaluated with laboratory momenta. And
baryon beams are likely more effective than pion beams in producing ccq and
cqq baryons at high X$_F$.

Consider a hadronic interaction in which two $c\bar c$ pairs are produced.
The two c's combine and then form a ccq baryon. Calculations for ccq
production via such interactions have not yet been published. Even if they
are done, they will have large uncertainties. Some ingredients to the
needed calculations can be stated. For ccq production, one must produce two
c quarks (and associated antiquarks), and they must join to a tightly
bound, small size anti-triplet pair. The pair then joins a light quark to
produce the final ccq. The two c-quarks may arise from two parton showers
in the same h-h collision, or even from a single parton shower, or they may
be present as an intrinsic charm component of the incident hadron, or
otherwise. The two c-quarks may be produced (initial state) with a range of
relative momenta and separations. In the final state, they are tightly bound
in a very small size cc pair, with low relative momentum. The overlap
integral between initial and final states determines the probability for
the cc-q fusion process. For cqq production, a produced c quark may more
easily combine with a (projectile) di-quark to produce a charmed baryon. A
ccq production calculation in this framework, based on two parton showers
in the same h-h collision, is in progress by Levin \cite {lev}.

As an aid in comparing different possible calculations, one may parametrize
the yield as:
\begin{equation}
\sigma(ccq)/\sigma(cqq) \approx \sigma(ccq)/\sigma(cq)
\approx k[\sigma(c\bar{c})/\sigma(in.)] \approx kR.
\end{equation}
Here, $\sigma(c\bar{c})$ is the charm production cross section, roughly 25
$\mu$b; $\sigma(in.)$ is the inelastic scattering cross section, roughly 25
mb; and R is their ratio, roughly 10$^{-3}$ \cite{cse}. Here, k is the
assumed "suppression" factor for joining two c's together with  a third
light quark to produce ccq; compared to cqq or cq production, where the c
quark combines with a light diquark to give cqq or a light quark to give
cq. Eq. 2 does not represent a calculation, and has no compelling
theoretical basis. It implicitly factorizes ccq production into a factor
(R) that accounts for the production of a second c-quark, and a factor (k)
describing a subsequent ccq baryon formation probability. Considering the
overlap integral described in the preceding paragraph, one may expect k
values less than unity for simple mechanisms of ccq formation. It is
possible to have a factor k$>$1, if there is some enhancement correlation
in the production mechanism. Reliable theoretical cross section
calculations are needed, including the X$_F$-dependence of ccq production.
In the absence of such a calculation, we will explore the experimental
consequences of a ccq search for the range k=0.1-1.0, corresponding to
$\sigma(ccq)/\sigma(cqq) \approx \sigma(ccq)/\sigma(cq) \approx
10^{-4}-10^{-3}$. Assuming $\sigma(c\bar{c})$ charm production cross
sections of 25 microbarns, this range corresponds to ccq cross sections of
2.5-25. nb/N.

Aoki et al. \cite {aok} reported a low statistics measurement at $\sqrt{s}$
=26 GeV for double to single open charm pair production, of $10^{-2}$. This
$D\bar{D}D\bar{D}$ to $D\bar{D}$ ratio was for all  central and diffractive
events. This high ratio is encouraging for ccq searches, compared to the
value from NA3 \cite {na3} of $\sigma{(\Psi\Psi)}/\sigma{(\Psi)} \approx 3
\times 10^{-4}$. We assume that the $\Psi\Psi$ result is relevant, even
though $\Psi$ production is only a small part ($\approx$ 0.4\%) of the
charm production cross section, with most of the cross section leading to
open charm. For double $\Psi$ or double charm pair hadroproduction, the
suppression factor k for two c-quarks to join into the same ccq is missing.
These two results for double charm production therefore establish a range
of values for R in Eq. 2, consistent with the value 10$^{-3}$ estimated
above in the discussion of Eq. 2. Robinett \cite {rob} discussed $\Psi\Psi$
production and Levin \cite {lev} discussed ccq production in terms of
multiple parton interactions. Halzen et al. \cite {hal} discussed evidence
for multiple parton interactions in a single hadron collision, from data on
the production of two lepton pairs in Drell-Yan experiments.

It will be of interest to compare ccq production in hadron versus electron-
positron collisions, even if CHEOPS deals with hadron interactions.
Following production of a single heavy quark from the decay of a Z or W
boson produced in an electron-positron collision, Savage and Wise \cite
{sw} discussed the expected suppression for the the production of a second
heavy quark by string breaking effects or via a hard gluon. Kisselev et al.
\cite {kis} calculated low cross sections for double charm production at an
electron-positron collider B factory, for $\sqrt{s}$= 10.6 GeV. They find
$\sigma(ccq)/\sigma(c\bar{c}) = 7. \times 10^{-5}$. Although this result is
inapplicable to hadronic interactions as in CHEOPS; the  work describes
some important calculational steps, and also demonstrates the continued
wide interest in this subject. Kisselev et al. \cite{kis2} give a
preliminary estimate of $\sigma(ccq) \approx 10. $ nb/N in hadronic
production at $\sqrt{s}$= 100. GeV.  This corresponds to k=0.4 in the
parametrization of Eq. 2.

A number of works \cite {atlas,fm,lo,ws,sl1} consider the production and
decay of doubly heavy hadrons (bcq, $\bar{b}c$, etc.) at future hadron
collider experiments at the FNAL Tevatron or CERN LHC. For example, the
process $gg \rightarrow  b \bar{b}$ or $q\bar{q} \rightarrow  b \bar{b}$
may be followed by gluon bremstrahlung and splitting $\bar{b} \rightarrow
\bar{b}g \rightarrow \bar{b}c\bar{c}$ (fragmentation) to yield B$_c$
($\bar{b}c$) mesons \cite {fal,ss,nas,bcy,chang,lmp,gls,sl2}. Doubly
charmed baryon production may then possibly proceed via the weak decay $b
\rightarrow c\bar{c}s$. This quark process has recently been claimed \cite
{dun} to dominate charm baryon production in B decay. A CHEOPS fixed target
study for ccq (possibly including some B$_c$ mesons) can be a valuable
prelude to collider studies of doubly heavy hadrons.

   Brodsky and Vogt \cite {bro} suggested that there may be significant
$c\bar{c}$ components in hadron wave functions at large X$_F$, and
therefore also $cc\bar{c}\bar{c}$ components. The double intrinsic charm
component can lead to ccq production, as the cc pairs pre-exist with as
many as one-third in the required anti-triplet combination, and they can be
released in hard processes. Brodsky and Vogt \cite {bro} discussed double
$\Psi\Psi$ production \cite {na3} in the framework of intrinsic charm. They
claim that the data (transverse momentum, $X_F$ distribution, etc.)
suggests that $\Psi\Psi$ production is highly correlated, as expected in
the intrinsic charm picture. A recent experiment \cite {kod} searched for
diffractive production of open charm in $D\bar{D}$ pairs with a 800 GeV
proton beam. The data were used to set a 0.2\% upper limit for the
intrinsic charm component for the proton wave function. A ccq production
estimate in the intrinsic charm framework would be of interest.

We can also refer to an empirical formula which reasonably describes the
production cross section of a mass M hadron in central collisions. The
transverse momentum distribution at not too large p$_t$ follows a form
given as \cite {hag}:
\begin{equation}
d\sigma/dp_t^2 \sim exp(-B\sqrt{M^2+p_t^2}),
\end{equation}
where B is roughly a universal constant $\sim$ 5 - 6 (GeV)$^{-1}$. The
exponential (Boltzmann) dependence on the transverse energy $E_t =
\sqrt{M^2+p_t^2}$ has inspired speculation that particle production is
thermal, at a temperature B$^{-1}$ $\sim$ 160 ~MeV \cite {hag}. We assume
that this equation is applicable to ccq production. To illustrate the
universality of B, we evaluate it for a few cases. For $\Lambda_c$ and
$\Xi^0$, empirical fits to data give exp(-$bp_t^2$), with b=1.1 GeV$^{-2}
$and b=2.0 GeV$^{-2}$, respectively \cite{wa89,rot}. With B $\approx$ 2Mb,
this corresponds to B=~5.0 GeV$^{-1}$ for $\Lambda_c$, and B=~5.3
GeV$^{-1}$ for $\Xi^0$. For inclusive pion production, experiment gives
exp(-$bp_t$) with b =~6 GeV$^{-1}$ \cite {pi6}; and B $\sim$ b, since the
pion mass is small. Therefore, B= 5-6 GeV$^{-1}$ is valid for $\Lambda_c$,
$\Xi^0$ hyperon, and pion production. After integrating over p$_t^2$,
including a (2J+1) statistical factor to account for the spin of the
produced ccq, and taking the mass of ccq and D to be 3.7 and 2.0 GeV
respectively; we estimate the ratio as:
\begin{equation}
\sigma(ccq)/\sigma(D) \sim (2J+1)exp[-5[M(ccq)-M(D)]]
\sim 4 \times 10^{-4}.
\end{equation}
This result corresponds to k=0.4 in the paramtrization of Eq. 2. In
applying Eq. 4 to ccq production, we assume that the suppression of cross
section for the heavy ccq production (for q = u,d,s) as compared to the
light D ($\bar{c}q$) production is due to the increased mass of ccq.
However, this formula ignores important dynamical input, including
threshold effects and a possible suppression factor for the extra charm
production in ccq, and therefore can be considered an upper limit. One may
apply Eq. 4 with appropriate masses to estimate yield ratios of other
particles. For the T, we assume the same production cross section as for
the ccq, based on the mass dependence of Eq. 4.

\section*{Decay Modes and Branching Ratios of ccq Baryons}

The semileptonic and nonleptonic branching ratios of ccq baryons have been
estimated by Bjorken \cite {bj} in unpublished notes of 1986. He uses a
statistical approach to assign probabilities to different decay modes. He
first considers the most significant particles in a decay, those that carry
baryon or strangeness number. Pions are then added according to a Poisson
distribution. The Bjorken method and other approaches for charm baryon
decay modes are described by Klein \cite {kle}. Savage and Springer \cite
{ss} examined the flavor SU(3) predictions for the semileptonic and
nonleptonic ccq weak decays. They give tables of expected decay modes,
where the rates for different modes are given in terms of a few reduced
matrix elements of the effective hamiltonian. In this way, they also find
many relationships between decay rates of different modes. Savage and
Springer discuss the fact that the SU(3) predictions for the decay of the
D-mesons can be understood only by including the effects of final state
interactions \cite {cc}. They suggest that FSI effects should be much less
important for very charming baryons (ccq) compared to charmed mesons.

The c decays weakly, for example by $c \rightarrow s + u\bar{d} +
n(\pi^+\pi^-)$, with n=0,1, etc. In that case, for example, $ccs
\rightarrow css$ + ($\pi^+ \pi^+ \pi^-$ or $\rho^+ \pi^+ \pi^-$). The event
topology contains two secondary vertices. In the first, a css baryon and 3
mesons are produced. This vertex may be distinguished from the primary
vertex, if the ccs lifetime is sufficiently long. The css baryon now
propagates some distance, and decays at the next vertex, in the standard
modes for a css baryon. The experiment must identify the two secondary
vertices.

We describe some decay chains considered by Bjorken \cite {bj}. For the
$\Xi_{cc}^{++}$, one may have $\Xi_{cc}^{++} \rightarrow $ $\Sigma_c^{++}
K^{*0} $ followed by $\Sigma_c^{++} \rightarrow \Lambda_c^{+} \pi^{+}$ and
K$^{*0} \rightarrow  K^{-} \pi^{+}$. A $\Lambda_c^{+} \pi^{+}   K^{-}
\pi^{+}$ final state was estimated by Bjorken \cite {bj} to have as much as
5\% branching ratio. Bjorken also estimated a 1.5\% branch for
$\Xi_{cc}^{++} \rightarrow $ $\Xi_c^{+} \pi^+$; and 1.5\% for
$\Omega_{cc}^+ \rightarrow \Xi_c^+ \pi^+ K^-$. Bjorken finds that roughly
60\% of the ccq decays are hadronic, with as many as one-third of these
leading to final states with all charged hadrons. The decay topologies
should satisfy a suitable CHEOPS charm trigger, with reasonable efficiency.
There are also predicted 40\% semi-leptonic decays. However, with a
neutrino in the final state, it is not feasible to obtain the mass
resolution required for a double charm search experiment.

\section*{Decay Modes and Branching Ratios of the T}
One can search for the decay of T $\rightarrow \pi$ D D, or T $\rightarrow
\gamma$ D D, as discussed by Nussinov \cite {nus}. The pion or gamma are
emitted at the primary interaction point, where the D* decays immediately.
The two D mesons decay downstream. The D* decay to $\pi$-D is more useful
for a search, since the charged pion momentum can be  measured very well.
One can then get very good resolution for the reconstruction of the T mass.
For the gamma decay channel, the experimental resolution is worse. There
will therefore be relatively more background in this channel, since the
gamma multiplicity from the target is high, and one must reconstruct events
having two D mesons, with all gammas.

\newpage
\section*{Signal and Background Considerations}

High energies are needed for studies of high mass, and short lifetime
baryons. Thereby, one produces high energy doubly charmed baryons. The
resulting large lifetime boost improves separation of secondary and primary
vertices, and improves track and event reconstruction. CHEOPS with 450 GeV
protons or other 350-450 GeV hadrons \cite {paul} has this high energy
advantage.

One  can identify charm candidates by requiring that one or more decay
particles from a short lived parent have a sufficiently large impact
parameter or transverse miss distance relative to the primary interaction
point. This transverse miss distance (S) is obtained via extrapolation of
tracks that are measured with a high resolution detector close to the
target. This quantity is a quasi-Lorentz invariant. Consider a relativistic
unpolarized parent baryon or a spin zero meson that decays into a daughter
that is relativistic in the parent's center of mass frame. Cooper \cite
{pc1} has shown that the average transverse miss distance is $S \approx \pi
c \tau/2$. For example, $\Lambda_c$ with c$\tau$ $\approx$ 60 microns
should have S $\approx$  90 microns. The E781 on-line filter cut is on the
sum of the charged decay products of the doubly charmed baryon and the
singly charmed baryon daughter's decay products. Any one of these with
P$>$15 GeV/c and S$>$30 microns generates a trigger \cite {russ}. Events
from the primary vertex are typically rejected by the cut on S. With a
vertex detector with 20 micron strips, the E781 resolution in S is about 4
microns for very high momenta tracks. For events in E781 with a 15 GeV
track, the transverse miss-distance resolution deteriorates to about 9
microns, due to multiple scattering \cite {pc2}. And the resolution gets
even worse for yet lower momenta tracks. As this resolution becomes worse,
backgrounds increase, since the S-cut no longer adequately separates charm
events from the primary interaction events. The backgrounds are not only
events from the primary vertex, but also from the decays of the hadrons
associated with the two associated $\bar{c}$ quarks produced together with
the two c quarks. It is expected that the requirement to see two related
secondary vertices may provide a significant reduction in background
levels.

Some bqq production and decay, with two secondary vertices, may be observed
in CHEOPS, and must be considered at least as background to ccq production.
The bqq and ccq events may be distinguished by the larger bqq lifetime, and
the higher transverse energy released in the b decay. It is not the aim of
CHEOPS to study bqq baryons. Experiments at CERN gave only a small number of
reconstructed bqq baryons, at a center of mass energy around 30 GeV \cite
{cernb}.

CHEOPS considers using a  multiplicity jump trigger \cite{kwa}, which is
intended to be sensitive to an increase in the number of charged tracks
following a charm decay. Such a trigger for high rate beams has not yet
been used in a complete experiment, and still requires research and
development. Backgrounds are possible with such a trigger, due to secondary
interactions in targets and the interaction detector (Cerenkov, possibly
\cite {paul}) following each target. Also, gamma rays from a primary
interaction may convert afterwards to electron-positron  pairs, and falsely
fire the trigger. If the rejection ratio of such non-charmed events is not
sufficiently high, the trigger may not achieve its needed purpose of
reducing the accepted event rate to manageable values. This trigger would
be sensitive to events with X$_F$ $>$ -.1, and therefore has effectively an
"open" trigger X$_F$-acceptance. Most of the charm events accepted will
then be mainly associated with charm mesons near X$_F$=0, since these
dominate the  cross section in hadronic processes. The decay of ccq to a
singly charmed hadron may trigger, or the charmed hadron's decay may fire
the trigger. The event also has two anticharmed quarks, associated with
charmed hadrons, and they may also fire the trigger. However, low-X$_F$
events may have high backgrounds, since it is more difficult to separate
them from non-charmed events, due to the poor miss distance resolution. For
higher X$_F$  events, one obtains a sample of doubly charmed baryons with
improved reconstruction probability because of kinematic focussing and
lessened multiple scattering and improved particle identification. The
multiplicity-jump trigger for CHEOPS could be supplemented by a momentum
condition trigger P $>$ 15 GeV/c, similar to this requirement in E781. This
could enhance the high-X$_F$ acceptance, and give higher quality events.

For double charm, the target design is important. To achieve a high
interaction rate and still have small multiple scattering effects, one may
choose five 400 micron Copper targets, separated by 1 mm. The total target
thickness is limited to 2\% interaction length in order to keep multiple
scattering under control. With different target segments, one requires a
longitudinal tracking resolution of 200-300 microns, in order to identify
the target segment associated with a given interaction. The knowledge of
the target segment allows the on-line processor to reconstruct tracks, and
identify a charm event. The tracking detectors would then be placed as
close as possible to the targets, to achieve the best possible transverse
miss-distance resolution. The optimum target design and thickness for
double charm requires study via Monte Carlo simulation.

One may require separation distances of secondary from primary vertices of
$\approx$ 1-4 $\sigma$, depending on the backgrounds. The requirement for
two charm vertices in ccq decays may reduce backgrounds sufficiently, so
that this separation distance cut is less important than in the case of cqq
studies.  For a lifetime of $~100fs$, with a laboratory lifetime boost of
15, the distance from the production point to the decay point is around 450
microns. E781 can attain roughly 300 micron beam-direction resolution for
X$_F$ =0.2, with a 650 GeV beam, and 20 micron strip silicon detectors. For
lower X$_F$  events, the resolution deteriorates due to multiple
scattering, and there is little gain in using narrower strips. CHEOPS aims
to achieve 150 micron resolution for the high X$_F$  events.
Signal and background and trigger simulations and target design development
work are in progress for  CHEOPS \cite {paul}.

\section*{Projected Yields for CERN CHEOPS}

For CHEOPS with a Baryon beam, one may rely on previous measurements done
with similar beams. The open charm production cross section at SPS energies
is roughly 25 $\mu$b. Taking Eq. 2 with a reduction factor of kR=4.
$\times$ 10$^{-4}$, with k=0.4, we have $\sigma$(ccq) $\approx 10.$ nb/N.
This kR value follows from Kiselev et al. \cite {kis2} and from Eq. 4. We
assume a measured branching ratio B= 10\% for the sum of all ccq decays;
this being 50\% of all the decays leading to only charged particles. We
also assume a measured B = 20\% for the sum of all cqq decays, this being
roughly the value achieved in previous experiments. With these branching
ratios, we estimate $\sigma \cdot BB = 10. \times 0.2 \times 0.1 = 0.2
nb/N$.

For CHEOPS,  we now evaluate the rate of reconstructed ccq events. The
expectations are based on a beam of 5. $\times 10^7$ per spill, assuming
240 spills per hour of effective beam, or 1.2 $\times 10^{10}$/hour. For a
4000 hour run (2 years), and a 2\% interaction target, one achieves 9.5
$\times 10^{11}$ interactions per target nucleon. We assume that
$\sigma$(charm) = 25 $\mu$b and $\sigma$(in) = 25 mb for a proton target,
and take a charm production enhancement per nucleon of A$^{1/3}$ (with mass
A $\approx$ 64 for CHEOPS). One then obtains a high sensitivity of 1.5
$\times 10^{5}$ charm events for each nb per nucleon of effective cross
section (for nucleons in A $\approx$ 64 nuclei), where $\sigma_{eff} =
\sigma BB \varepsilon$. Here $\varepsilon$ is the overall efficiency for
the experiment. Fermilab E781 with  650 GeV pion and $\Sigma^-$ beams is
scheduled for 1996-97. This experiment may therefore observe ccq baryons
before CHEOPS, as described in recent reports \cite{mmi,gru}. The CHEOPS
Letter of Intent \cite {paul} describes plans to achieve roughly ten times
more reconstructed charm events than Fermilab E781. However, a data run for
CHEOPS is not yet scheduled.

We consider also the expected CHEOPS efficiency for the charm events, by
comparison to E781 estimated \cite {russ} efficiencies. The E781
efficiencies for cqq decays include a tracking efficiency of 96\% per
track, a trigger efficiency averaged over X$_F$ of roughly 18\%, and a
signal reconstruction efficiency of roughly 50\%. The CHEOPS trigger
efficiency for cqq should be higher than E781, if low X$_F$ events are
included. However, the signal reconstruction efficiency is low for low
X$_F$ events. The reconstruction efficiency should be lower for double
charm events, since they are more complex than single charm events. Yet,
using the proposed type of vertex detector, multivertex events can be
reconstructed with good efficiency \cite {cernb}. In a spectator decay
mode, the final state from ccq decay will likely be a csq charm baryon plus
a W decay, either semileptonic (40\% total B) or hadronic (25\% $\pi^+$,
75\% $\rho^+$ most likely). One may expect the vertex to be tagged more
often (roughly a factor of two) for double charm compared to single charm.
There should therefore be a higher trigger efficiency and a lower
reconstruction efficiency for double compared to single charm. We assume
here however that the product of these two efficiencies remains roughly the
same. Therefore, the overall average ccq efficiency is taken to be
$\varepsilon \simeq 8\%$, comparable to the expected E781 value for cqq
detection. The expected yield given above is 1.5 $\times 10^{5}$ charm
events/(nb/N) of effective cross section. For $\sigma$BB = 0.2 nb/N, one
has $\sigma_{eff} = 0.016 nb/N$, and therefore N(ccq) $\approx$  2400
events for  CHEOPS. This is the total expected yield for ccu,ccd,ccs
production for ground and excited states. For k $>$ 0.4, the yields are yet
higher.

\section*{Conclusions}
The observation of doubly charmed baryons or T would make possible a
determination of their lifetimes and other properties. The expected low
yields and short lifetimes make double charm hadron research an
experimental challenge. The discovery and subsequent study of the ccq
baryons or T should lead to a deeper understanding of the heavy quark
sector.

\newpage
\section*{Acknowledgements}

Thanks are due to  P. Cooper, F. Dropman, L. Frankfurt, S. Gavin, J.
Grunhaus, K. Konigsmann,  B. Kopeliovich, E. Levin, H. J. Lipkin, B.
Muller, S. Nussinov,  S. Paul, B. Povh, J. Russ, M. A. Sanchis-Lozano, M.
Savage, R. Werding, and M. Zavertiev  for stimulating discussions. This
work was supported in part by the U.S.-Israel Binational Science Foundation
(B.S.F.), Jerusalem, Israel.


\begin{thebibliography}{99}
\bibitem{bj} J. D. Bjorken, FERMILAB-CONF-85/69, Is the ccc a New Deal for
Baryon Spectroscopy?, Int. Conf. on Hadron Spectroscopy, College Park, MD,
Apr. 1985, Published in College Pk. Hadron. Spect., (QCD161:I407:1985, P.
390.);\\ Unpublished Draft, "Estimates of Decay Branching Ratios for
Hadrons Containing Charm and Bottom Quarks", July 22, 1986;\\ Unpublished
Draft, "Masses of Charm and Strange Baryons", Aug. 13, 1986.
\bibitem {ric} J.M. Richard, in Proc. of the CHARM2000 Workshop, see Ref.
\cite {kk}, HEPPH-9407224, (Bulletin Board: hep-ph@xxx.lanl.gov -
9407224),\\ E. Bagan, H. G. Dosch, P. Gosdzinsksy, S. Narison, J. M.
Richard,\\ Z. Phys. C64 (1994) 57,\\ S. Zouzou, J. M. Richard, Few-Body
Systems 16 (1994) 1,\\ HEPPH-9309303\\ J. M. Richard, Nucl. Phys. B 21
(Proc. Suppl.) (1991) 254.
\bibitem {fr} S. Fleck, J. M. Richard, Prog. Theor. Phys. 82 (1989) 760;\\
S. Fleck, J. M. Richard, Particle World 1 (1990) 67.
\bibitem{sw} M. J. Savage, M. B. Wise, Phys. Lett B248 (1990) 177.
\bibitem{ss} M. J. Savage, R. P. Springer, Int. J. Mod. Phys. A6 (1991) 1701.
\bibitem{kis} V. V. Kiselev et al., Phys. Lett. 332 (1994) 411;\\ V. V. Kiselev
et al., Preprint IHEP 94-10 (1994);\\ V. V. Kiselev et al., Yad. Fiz. 57 (1994)
733.
\bibitem {kis2} V. V. Kiselev et al., Sov. J. Nucl. Phys. 46 (1987) 535.
\bibitem {fal} A. F. Falk, M. Luke, M. J. Savage, M. B. Wise,
Phys. Rev. D49 (1994) 555.
\bibitem {ban} M. Bander, A. Subbaraman, Phys. Rev. D50 (1994) 5478.
\bibitem{app} J. A. Appel, Ann. Rev.  Nucl. Part.  Sci.  42 (1992) 367.
\bibitem {tav} S. P. K. Tavernier, Rep. Prog. Phys. 50 (1987) 1439.
\bibitem {kle} S. R. Klein, Int. Jour. Mod. Phys. A5 (1990) 1457.
\bibitem {pdg} Particle Data Group, Phys. Rev. D50 (1994) 1171.
\bibitem {sie} H. W. Siebert, Nucl. Phys. B21 (Proc. Suppl.)
(1991) 223.
\bibitem{kor} J. G. Korner, H. W. Siebert, Ann. Rev. Nucl. Part.
Sci., 41 (1992) 511.
\bibitem{paul} S. Paul et al., Letter of Intent,
CHEOPS, CHarm Experiment with Omni-Purpose Setup, CERN/SPSLC 95-22,
SPSLC/I202, March 28, 1995.
\bibitem {kk} D. M. Kaplan and S. Kwan, Editors, Proc. of the CHARM2000
Workshop, The Future of High Sensitivity Charm Experiments, Fermilab, June
1994, FERMILAB-CONF-94/190.
\bibitem {ros} J. L. Rosner, Enrico Fermi
Institute and U. Chicago preprint  EFI-95-02,  HEPPH-9501291, Jan. 1995.
\bibitem{lip} H. J. Lipkin,
Phys. Lett. B45 (1973) 267; Phys. Lett. 172B (1986) 242.
\bibitem {tor} N. A.
Tornqvist, Z. Phys. C61 (1994) 525; Phys. Rev. Lett. 67 (1991) 556;
Il Nuovo Cimento 107A (1994) 2471.
\bibitem{nus} S. Nussinov, Unpublished Draft, Private Communication, 1995.
\bibitem {cho} Chi-Keung Chow,
Preprint CALT-68-1964,  HEPPH-9412242, 1994.
\bibitem {bein} M. W. Beiner, B. C. Metsch, H. R. Petry, \\
HEPPH-9505215, submitted to Zeit. fur Phys. A, 1995.
\bibitem {lan1}
L. G.~Landsberg, Surveys in High Energy Phys.  6 (1992) 257;\\
K.~Peters, Proceed. of LEAP-92 Conf., Courmayeur, Aosta Valley, p.93, Sep.
14-19, 1992, Eds. C.~Guaraldo et al., 1993, North-Holland;\\
L. G.~Landsberg, Yad.Fiz. 57 (1994) 47;\\
L. G. Landsberg, M. A. Moinester, M. A. Kubantsev,\\
Report IHEP 94-19, TAUP 2153-94, Protvino and Tel Aviv, 1994.
\bibitem {moi} M. A. Moinester, D. Ashery, L. G. Landsberg, H. J. Lipkin,\\
in Proc. CHARM2000 Workshop, ibid., Fermilab, June 1994,\\
Tel Aviv U. Preprint TAUP 2179-94, HEPPH-9407319.
\bibitem {zie} M. Zielinsky et al., Z. Phys. C31 (1986) 545; C34 (1987)
255.
\bibitem {u3100} K. Martens, Dr. Rer. Nat. thesis, U. Heidelberg,\\
"Die Suche nach dem Zerfall U$^+$(3100) $\rightarrow \Lambda \bar{p} \pi^+
\pi^+$ in dem\\ Hyperonenstrahlexperiment WA89", Nov. 1994.
\bibitem {H} M. A. Moinester, C. B.
Dover, H. J. Lipkin, Phys. Rev. C46 (1992) 1082.
\bibitem{lif1} G. Bellini, Invited
talk, Int. Workshop "Heavy Quarks in Fixed Target", Charlottesville, Virginia,
Oct. 1994.
\bibitem {lif2} M. B. Voloshin, M. A. Shifman, Sov. Phys. JETP 64
(1986) 698.
\bibitem {lif3} R. Forty, "Lifetimes of Heavy Flavor Particles",
CERN-PPE/94-144, Sep. 1994, Invited talk, XIV Int. Conf. on Physics in
Collision,\\ Tallahassee, Florida, June 1994.
\bibitem {fra} L. Frankfurt, Private Communication, 1994.
\bibitem{lev} E. Levin, Private Communication, 1995.
\bibitem{cse} B. D'Almagne, Int.  Symposium
on the Production and Decay of Heavy Flavors,  Stanford, Cal., Sep. 1987,
SLAC Heavy Flavors 1987 (QCD161:I732:1987); \\
K. Kodama et al., E653 Coll., Phys. Lett. B263
(1991) 579;\\
M. L. Mangano, P. Nason, G. Ridolfi, Nucl. Phys. B405 (1993)
507.
\bibitem{aok} S. Aoki et al., CERN WA75 collaboration, Phys. Lett. B187 (1987)
185.
\bibitem{na3} J. Badier et al., CERN NA3 collaboration,
Phys. Lett. B158 (1985) 85, Phys. Lett. B114 (1982) 457, Phys. Lett. B124 (
1983) 535.
\bibitem{rob} R. W. Robinett, Phys. Lett. B230 (1989) 153.
\bibitem{hal} F. Halzen, P. Hoyer, W. Y. Stirling, Phys. Lett. B188 (1987)
375.
\bibitem{atlas} ATLAS internal note, Phys-NO-041 (1994);\\ ATLAS internal
note Phys-NO-058 (1994).
\bibitem{fm} A. Fridman, B. Margolis, CERN preprint CERN-TH 6878/93
(1993).
\bibitem{lo} P. Lebrun, R.J. Oakes, Preprint FERMILAB-Conf-93/303
(1993).
\bibitem{ws} M. J. White, M. J. Savage, Phys. Lett. B271 (1991) 410.
\bibitem{sl1} M. A. Sanchis-Lozano, Phys. Lett. B321 (1994) 407.
\bibitem{nas} P. Nason et al., CERN preprint CERN-TH.7134/94 (1994).
\bibitem {bcy} E. Braaten, K. Cheung, T.C. Yuan, Phys. Rev. D48 (1993) 5049.
\bibitem {chang} C-H Chang, Y-Q Chen, Phys. Rev.  D48 (1993) 4086.
\bibitem {lmp} M. Lusignoli, M. Masetti, S. Petrarca,
Phys. Lett. B266 (1991) 142.
\bibitem {gls} S. S. Gershtein, A.K. Likhoded, S.R. Slabospitsky,
Int. J. Mod. Phys. A13 (1991) 2309.
\bibitem{sl2} M. A. Sanchis-Lozano, submitted to Nucl. Phys. B,
HEPPH-9502359, Mar. 15, 1995.
\bibitem{dun} I. Dunietz, P. S. Cooper, A, F. Falk, M. B. Wise, \\
Phys. Rev. Lett. 73 (1994) 1075.
\bibitem{bro} R. Vogt, S. J. Brodsky, SLAC-PUB-95-6753,
LBL-36754, HEPPH-9503206, Feb. 1995, Submitted to
Phys. Lett. B; R. Vogt, S. J. Brodsky, SLAC-PUB-6468, LBL-35380,
HEPPH-9405236, Apr. 1994,\\ R.V. Gavai, S. Gupta, P.L. McGaughey, E. Quack,
P.V.
Ruuskanen, R. Vogt, Xin-Nian Wang, GSI-94-76,  HEPPH-9411438, Nov. 1994,\\
R. Vogt, Nucl. Phys. A553 (1993) 791c,\\ R. Vogt, S. J. Brodsky, P. Hoyer,
Nucl. Phys. B383 (1992) 643,\\ R. Vogt, S. J. Brodsky, P. Hoyer, Nucl.
Phys. B360 (1991) 67,\\ R. Vogt, LBL-37655, HEPPH-9503207, Jan. 1995.
\bibitem {kod} K. Kodma et al., Phys. Lett. B316 (1993) 188.
\bibitem {hag} R. Hagedorn, "The Long Way to the Statistical Bootstrap
Model",\\ CERN-TH-7190-94, Mar. 1994;\\
R. Hagedorn, in Quark Matter 84, ed. K. Kajantie,
Lecture Notes in Physics Vol. 221 (Springer-Verlag, New York, 1985); \\
H. Grote, R. Hagedorn, J. Ranft, "Atlas of Particle  Production Spectra", CERN
Report, 1970.
\bibitem {wa89} A. Simon, CERN WA89 Moriond Report,
Mar. 1994.
\bibitem {rot} F. S. Rotondo, Phys. Rev. D47, (1993) 3871.
\bibitem{pi6} S. D. Ellis and R. Stroynowski, Rev. Mod. Phys. 49 (1977) 753.
\bibitem {cc} L. Chau, H. Cheng, Phys. Rev. D36 (1987) 137, Phys. Lett.
B222 (1989) 285;\\
A. Kamal, R. Verma, Phys. Rev. D35 (1987) 3515, D36 (1987) 3527 (E).
\bibitem{pc1} P. Cooper, "Miss Distance Distribution", unpublished report,\\
Aug. 5, 1981, private communication.
\bibitem{russ} J. Russ, in Proc. CHARM2000 Workshop, ibid.,
Fermilab, June 1994.
\bibitem{pc2} J. Russ, E781 Internal Report, Oct. 29, 1993;\\
P. Cooper, E781 Internal Report, Feb. 27, 1994;\\ "Multiple Coulomb
Scattering in Silicon Tracking".
\bibitem{cernb} D. Barberis, CERN
WA92 BEATRICE collaboration, \\in Proc.
CHARM2000 Workshop, ibid., Fermilab, June 1994.
\bibitem{kwa} A. M. Halling, S. Kwan, Nucl. Inst. Meth. A333 (1993) 324.
\bibitem{mmi} M. A. Moinester, Tel Aviv U. Internal Report, Oct. 1994.
\bibitem{gru} J. Grunhaus, P. Cooper, J. Russ,\\
Fermilab E781 Report H718, Dec. 1994.
\end{thebibliography}
\end{document}